# Shadowing in Inelastic Scattering of Muons on Carbon, Calcium and Lead at Low $x_{Bj}$.

M.R.Adams[6], S.Aïd[10,a], P.L.Anthony[9,b], D.A.Averill[6], M.D.Baker[11], B.R.Baller[4], A.Banerjee[15], A.A.Bhatti[16,c], U.Bratzler[16], H.M.Braun[17], H.Breidung[17], W.Busza[11], T.J.Carroll[12], H.L.Clark[14,d], J.M.Conrad[5,e], R.Davisson[16], I.Derado[12], S.K.Dhawan[18], F.S.Dietrich[9], W.Dougherty[16], T.Dreyer[1], V.Eckardt[12], U.Ecker[17,f], M.Erdmann[1,g], G.Y.Fang[5,h], J.Figiel[8], R.W.Finlay[14], H.J.Gebauer[12], D.F.Geesaman[2], K.A.Griffioen[15,i], R.S.Guo[6,j], J.Haas[1], C.Halliwell[6], D.Hantke[12,k], K.H.Hicks[14], V.W.Hughes[18], H.E.Jackson[2], D.E.Jaffe[6,l], G.Jancso[12,7], D.M.Jansen[16,m], Z.Jin[16], S.Kaufman[2], R.D.Kennedy[3,n], E.R.Kinney[2,o], T.Kirk[2,p], H.G.E.Kobrak[3], A.V.Kotwal[5], S.Kunori[10], J.J.Lord[16], H.J.Lubatti[16], D.McLeod[6], P.Madden[3], S.Magill[6,q], A.Manz[12], H.Melanson[4], D.G.Michael[5,r], H.E.Montgomery[4], J.G.Morfin[4], R.B.Nickerson[5,s], J.Novak[12], S.O'Day[10,t], K.Olkiewicz[8], L.Osborne[11], R.Otten[17], V.Papavassiliou[2,u], B.Pawlik[8], F.M.Pipkin† [5], D.H.Potterveld[2], E.J.Ramberg[10], A.Röser[17,v], J.J.Ryan[11], C.W.Salgado[4,w], A.Salvarani[3,x], H.Schellman[13], M.Schmitt[5,y], N.Schmitz[12], K.P.Schüler[18,z], G.Siegert[1,aa], A.Skuja[10], G.A.Snow[10], S.Söldner-Rembold[12,bb], P.Spentzouris[13,e], H.E.Stier† [1], P.Stopa[8], R.A.Swanson[3], H.Venkataramania[18], M.Wilhelm[1,cc], Richard Wilson[5], W.Wittek[12], S.A.Wolbers[4], A.Zghiche[2], T.Zhao [16]

(Fermilab E665 Collaboration)

[1] *Albert-Ludwigs-Universität Freiburg i. Br.,Germany*
[2] *Argonne National Laboratory, Argonne, Illinois 60439*
[3] *University of California, San Diego, California 92093*
[4] *Fermi National Accelerator Laboratory, Batavia,Illinois 60510*
[5] *Harvard University, Cambridge, Massachusetts 02138*
[6] *University of Illinois, Chicago, Illinois 60607-7059*
[7] *KFKI Research Institute for Particle and Nuclear Physics of the Hungarian Academy of Sciences, H-1525 Budapest, Hungary*
[8] *Institute for Nuclear Physics, Krakow, Poland*
[9] *Lawrence Livermore National Laboratory, Livermore, California 94551*
[10] *University of Maryland, College Park, Maryland 20742*
[11] *Massachusetts Institute of Technology, Cambridge, Massachusetts 02139*
[12] *Max-Planck-Institut für Physik, Munich, Germany*
[13] *Northwestern University, Evanston, Illinois 60208*
[14] *Ohio University, Athens, Ohio 45701*
[15] *University of Pennsylvania, Philadelphia, Pennsylvania 19104*
[16] *University of Washington, Seattle, Washington 98195*
[17] *University of Wuppertal, Wuppertal, Germany*
[18] *Yale University, New Haven, Connecticut 06511*

Present addresses:
[†] deceased
[a] University of Hamburg, D-22603 Hamburg, Germany.
[b] SLAC, Stanford, CA 94309, USA.
[c] The Rockefeller University, New York NY 10021, USA.
[d] Texas A&M University, College Station, TX 77843.
[e] Columbia University, New York, NY 10027.
[f] Jenfelderstr. 147, D-22045 Hamburg, Germany.
[g] Heidelberg University, D-69120, Heidelberg Germany.
[h] Dept. of Medical Physics, University of Wisconsin, Madison, WI 53706.
[i] College of William and Mary, Williamsburg, VA 23187, USA.
[j] Institute of Physics, Academia Sinica, Nankang, Taipei Taiwan.
[k] GSF - Forschungszentrum fuer Umwelt und Gesundheit GmbH, 85764 Oberschleissheim, Germany.
[l] SCRI, Florida State University, Tallahassee, FL 32606, USA.
[m] LANL, Los Alamos, NM 87545, USA.
[n] Rutgers University, Piscataway, NJ 08855, USA.
[o] University of Colorado, Boulder, CO 80309, USA.
[p] Brookhaven National Laboratory, Upton, NY 11973, USA.
[q] Argonne National Laboratory, Argonne, IL 60439, USA.
[r] California Institute of Technology, Pasadena, CA 91125, USA.
[s] Oxford University, Oxford OX1 3RH, UK.
[t] Fermi National Accelerator Laboratory, Batavia, IL 60510, USA.
[u] Illinois Institute of Technology, Chicago, IL 60616, USA.
[v] Klinikum Barmen, Abt. Radiologie, D-42283 Wuppertal, Germany.
[w] CEBAF, Newport News, VA 23606, USA.
[x] A.T. & T., Bell Labs, 2000 North Naperville Road, Naperville, IL USA.
[y] University of Wisconsin, Madison, WI 53706, USA.
[z] DESY, D-22603 Hamburg, Germany.
[aa] University of Wuppertal, D-42119 Wuppertal, Germany.
[bb] Albert-Ludwigs-Universität Freiburg, 79104 Freiburg, Germany.
[cc] Hoffmann-LaRoche, CH-4002 Basel, Switzerland.

**Abstract**

Nuclear shadowing is observed in the per-nucleon cross-sections of positive muons on carbon, calcium and lead as compared to deuterium. The data were taken by Fermilab experiment E665 using inelastically scattered muons of mean incident momentum 470 GeV/c. Cross-section ratios are presented in the kinematic region $0.0001 < x_{Bj} < 0.56$ and $0.1 < Q^2 < 80$ GeV$^2$. The data are consistent with no significant $\nu$ or $Q^2$ dependence at fixed $x_{Bj}$. As $x_{Bj}$ decreases, the size of the shadowing effect, as well as its $A$ dependence, are found to approach the corresponding measurements in photoproduction.

# 1 Introduction

Since the discovery that the per-nucleon cross section for lepton scattering off heavy nuclear targets is "shadowed" relative to deuterium, several experiments have quantified the effect in various kinematic regions for different nuclei [1]. The shadowing effect had been observed in hadroproduction, photoproduction [2] and leptoproduction [3] experiments, but was thought to disappear with increasing four-momentum transfer squared, $Q^2$, when $x_{Bj} = Q^2/2M\nu$ was kept fixed [4]. This expected $Q^2$ dependence has not been observed, and several theoretical models have been developed in order to explain both the lack of $Q^2$ dependence and the observed $x_{Bj}$ dependence in the data.

The Vector Meson Dominance (VMD) model has had success in explaining the shadowing of real photons ($Q^2 = 0$), and several authors have extended this model to describe the interaction of virtual photons. In these models, the virtual photon is described in terms of a spectrum of vector mesons [5-7] or as a continuum of $q\overline{q}$ states [8, 9], which interact with a nucleon or the nucleus. Related to this type of interaction are the so-called "Large Rapidity Gap" events [10, 11], which result from diffractive scattering of the hadronic component of the photon [12, 13] and which have been previously observed by the E665 collaboration [14].

The virtual-photon nucleon interaction can also be viewed in an infinite-momentum frame. Then the scaling variable, $x_{Bj}$, is the fractional momentum of the quasi-free target parton, as long as $Q^2$ is "large" and $x_{Bj}$ remains finite. In this picture, a depletion of the per-nucleon cross section may be viewed as evidence that the parton distribution functions in heavy nuclear targets are in some way distorted relative to those of a free nucleon. This distortion may result from the interaction of partons from different nucleons, and therefore shadowing may reveal information on the long range nature of quantum chromodynamics [15-18]. Shadowing as measured in Drell-Yan leptoproduction [19], may be further evidence for the distortion of the nuclear structure functions by this mechanism.

In the lowest-order Born approximation, the double-differential cross section for inelastic lepton-nucleon scattering can be expressed as:

$$\frac{d^2\sigma}{dx_{Bj}dQ^2} = \frac{4\pi\alpha^2}{Q^4}\frac{F_2(x_{Bj},Q^2)}{x_{Bj}}\left[1 - y - \frac{Mx_{Bj}y}{2E} + \frac{y^2[1+4M^2x_{Bj}^2/Q^2]}{2[1+\sigma_l/\sigma_t]}\right] \tag{1}$$

where $Q^2$ is the negative square of the virtual-photon four-momentum, $M$ is the nucleon mass, $E$ is the laboratory energy of the incident muon, $\nu$ is the energy transferred to the hadronic system in the laboratory frame, $x_{Bj} = Q^2/2M\nu$, $y = \nu/E$, $\alpha$ is the electromagnetic coupling constant, $F_2$ is the structure function of the nucleon and $\sigma_l/\sigma_t$ is the ratio of the longitudinal to transverse photoabsorption cross sections. In this analysis [20], the ratios of the average per-nucleon cross-sections, $R^A(x_{Bj}) \equiv \frac{\sigma^A(x_{Bj})/A}{\sigma^D(x_{Bj})/2}$, also denoted by $A_{eff}/A$, are presented. Here $\sigma^A(x_{Bj})$ and $\sigma^D(x_{Bj})$ are the cross sections on a nucleus of atomic number $A$, and on deuterium respectively. Assuming $\sigma_l/\sigma_t$ to be independent of $A$ [21, 22], the cross-section ratios, $R^A(x_{Bj})$, are equal to the corresponding structure function ratios $F_2^A/F_2^D$. Measured ratios, $R^A(x_{Bj})$, are a function of $A$, $x_{Bj}$ and possibly of $Q^2$ or $\nu$. As $x_{Bj}$ approaches zero the cross-section ratios of heavy nuclear targets to deuterium are found to fall below unity. This depletion is known as "shadowing" and has been observed in various experiments [1, 23]. This paper reports measurements of virtual-photon shadowing for carbon, calcium and lead in the region $0.0001 < x_{Bj} < 0.56$ at $Q^2$ values between 0.1 and 80 GeV$^2$ and $\nu$ values between 50 and 300 GeV.

## 2 The Experiment

Details of the E665 spectrometer have been documented elsewhere [24]; only those components which are germane to this analysis will be discussed. The muon beam had a mean momentum of 470 GeV/c with a dispersion of 56 GeV/c. The momentum resolution of the beam spectrometer, $\delta p/p$, was typically 0.4%. In the 1990 run of E665, liquid targets of hydrogen and deuterium, and solid targets of carbon, calcium, and lead were interchanged approximately once per Tevatron cycle (58 seconds). Target cycling greatly reduced the systematic uncertainties due to time-dependent detector response. Target densities and incident muon fluxes are reported in Table 1. An open geometry spectrometer, instrumented with multiwire proportional chambers and drift chambers, determined the scattering angle and energy of the scattered muon upstream of 3 meters of steel, which served as a hadron absorber. The forward spectrometer provided a scattered-muon momentum resolution, $\delta p/p$, of better than 1.5%. Muons were identified by matching tracks reconstructed in the forward spectrometer with tracks found in four stations of proportional tubes and scintillating hodoscope planes located downstream of the steel. In addition, a gas sampling electromagnetic calorimeter, located in front of the steel absorber, had a resolution of $\delta E/E = (0.38 \pm 0.11)/\sqrt{E/\mathrm{GeV}}$ [20] for photons and electrons with energies below 80 GeV. The calorimeter was used to identify and remove background events.

| Target | Areal density (g/cm$^2$) | Number of interaction lengths | Number of radiation lengths | Number of scatters | Number of beam $\mu$'s ($\times 10^{10}$) |
|---|---|---|---|---|---|
| Hydrogen | 6.985$\pm$ 0.100 | 0.138 | 0.114 | 29968 | 6.07 |
| Deuterium | 15.880$\pm$0.100 | 0.292 | 0.131 | 37426 | 3.42 |
| Carbon(Thin) | 15.074$\pm$0.002 | 0.176 | 0.352 | 4713 | 0.51 |
| Carbon(Thick) | 29.952$\pm$0.005 | 0.349 | 0.699 | 33503 | 1.70 |
| Calcium | 19.4977$\pm$0.0007$^\dagger$ | 0.160 | 1.190 | 38131 | 3.12 |
| Lead | 5.370 $\pm$0.010 | 0.029 | 0.853 | 23114 | 7.40 |

$^\dagger$ See [25]

Table 1: Target properties, the observed number of scatters and beam flux after the cuts have been applied.

The use of a small-angle trigger (SAT) allowed E665 to trigger at scattering angles as small as 1 milliradian and thus to measure cross sections at $Q^2$ values down to 0.1 GeV$^2$. A relevant feature of the SAT was the use of a veto counter located upstream of the hadron absorber, in addition to the veto counters located behind the absorber. The upstream veto element reduced the rate of spurious triggers from muon scatters in the absorber, but also subjected the SAT to vetoes from final state hadrons. The systematic effects of such a veto are described below.

## 3 Selection of the Data

The following cuts were applied to all interactions reconstructed in the target region: $Q^2 > 0.1\,\mathrm{GeV}^2$ removed non-interacting beam muons, $\nu > 50$ GeV and $\delta x_{Bj}/x_{Bj} < 0.2$ excluded events with poor spectrometer resolution, and $y < 0.7$ eliminated poorly reconstructed scattered-muon tracks and the kinematic region where background processes were predominant. It was also required that exactly one muon, with momentum resolution better than 0.8%, was reconstructed in the beam spectrometer, and that beam and scattered muons satisfy a software simulation of the SAT.

In addition to these cuts, the electromagnetic calorimeter was used to identify and remove remaining backgrounds, coherent and quasi-elastic muon bremsstrahlung and elastic muon-electron scatters. If, in an event, the energy of the most energetic calorimeter cluster exceeded 21% of the muon's energy loss, $\nu$, the event was assumed to be muon-bremsstrahlung or elastic muon-electron scatter and was removed from the final sample. The number of inelastic scatters removed by such a cut was estimated as 17%.

Figure 1 shows the distribution of the energy of the most energetic calorimeter cluster divided by $\nu$ for those events which satisfy the resolution and kinematic selection. The distributions are shown for all scatters and for events defined as coherent muon bremsstrahlung scatters, elastic muon-electron scatters and inelastic scatters. These samples were defined based on event kinematics and final-state particle multiplicities. The muon bremsstrahlung scatters are required to have $Q^2 < 0.3$ GeV$^2$ and no charged particle other than the scattered muon reconstructed in the event. The muon-electron scatters have a single, reconstructed, negatively charged track with (track momentum/$\nu$) $> 0.75$ in addition to the scattered muon. The scatters of the last sample have at least two positively charged tracks or two negatively charged tracks reconstructed and fitted to the muon-muon scattering vertex. This requirement implicitly assumes that events which result in a high multiplicity final state are inelastic scatters. Figure 1 demonstrates that the cut in (Cluster Energy/$\nu$), indicated by the vertical line, very efficiently removes radiative backgrounds while retaining inelastic scatters.

Cross-section ratios were determined from the event samples after applying the cut on the calorimeter cluster energy. Alternatively, in order to enable a direct comparison with the shadowing results from the NMC collaboration [26], ratios were measured from the original event samples after applying radiative corrections as calculated by the FERRAD radiative correction program [26]. In the region where the radiative corrections are expected to be reliable, $y < 0.7$ and $x_{Bj} > 0.002$, the radiative corrected cross-section ratios are systematically lower than the ratios from the calorimeter analysis. For C/D they are lower by 2 to 6%, for Ca/D by 4 to 6% and for Pb/D by 4 to 8% (see Tables 4 to 6). It should be noted that these two methods of handling the radiative background are fundamentally different. The first one (calorimeter method) relies on eliminating the dominant $A$-dependent radiative background. In the second method, it is assumed that the radiative background contained in the raw data samples is well described by the analytical expressions (which are based on the calculations by Mo and Tsai [27]) used in the FERRAD program. The observed differences in the corrected cross-section ratios may indicate that these assumptions are not completely fullfilled. Because the FERRAD program is not reliable in the low-$Q^2$ region, and can also not be applied in the low-$x_{Bj}$ low-$Q^2$ region ($x_{Bj} < 0.001$, $Q^2 < 0.5\,\mathrm{GeV}^2$), this region being dominated by elastic muon-electron scattering, results will be presented using the calorimeter corrected data. In Tables 4 to 6 the FERRAD corrected ratios are given for comparison.

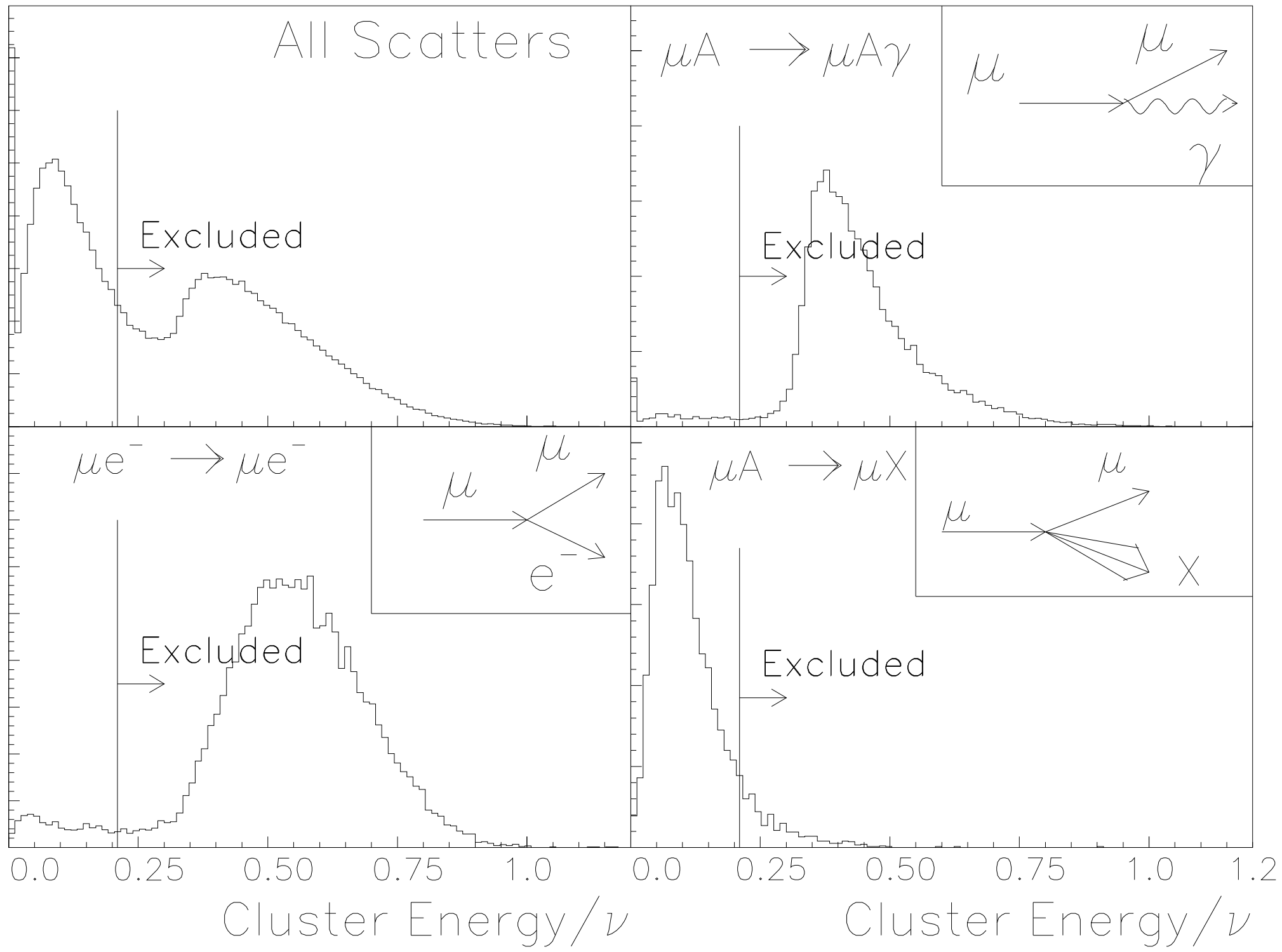

Figure 1: Distributions of the normalized energy of the most energetic calorimeter cluster in an event, (Cluster Energy)/$\nu$. The four samples represent all scatters and events classified as: muon-bremsstrahlung events, elastic muon-electron scatters and inelastic scatters. The vertical lines indicate the position of the cut in (Cluster Energy)/$\nu$. The vertical scale is arbitrary.

# 4 Corrections and Systematic Uncertainties

The uncorrected cross-section ratios were calculated from the raw event yields using

$$R^A(x_{Bj}) = \frac{w^A(x_{Bj})N^A(x_{Bj})/(d^A F_\mu^A)}{w^D(x_{Bj})N^D(x_{Bj})/(d^D F_\mu^D)} \ , \tag{2}$$

where $d^A$, $d^D$ and $F_\mu^A$, $F_\mu^D$ are the areal target densities and incident muon fluxes for target $A$ and deuterium respectively. $N^A$, $N^D$ are the number of scatters, and $w^A$, $w^D$ are weights assigned to each scatter to correct for each target's neutron excess. The weighting was based on the $F_2^n/F_2^p$ parameterization by the BCDMS collaboration [28]. The correction is only relevant for the Pb/D ratio, and the resulting correction was less than 1% for $x_{Bj} < 0.1$. A correction of $-1.01\%$ was applied to the deuterium target density to account for the presence of a 4.4% (per unit volume) contamination of HD. Finally, ratios were corrected bin-by-bin in $x_{Bj}$ to account for empty-target interactions. The magnitude of the empty-target correction was as large as 9% for the lead target and was consistent with that expected due to the amount of stray material in the beam to within 1% for all targets.

The beam flux was measured by counting a prescaled number of incident muons which satisfied the same "beam" requirement as a reconstructed scatter. Beam flux measured in this manner was found to be consistent, to better than 1%, with an independent measurement using scalers. The beam reconstruction was found to account for approximately 1% uncertainty. Temperature dependent fluctuations in the deuterium density have been measured and found to be less than 1%. An additional 0.63% uncertainty in luminosity was introduced by the shape of the cryogenic deuterium vessel which was rounded on the ends. The quoted target density and target dependent beam flux uncertainties were added quadratically to yield an overall systematic uncertainty in normalization of 1.30%, 1.33% and 1.55% on the carbon, calcium and lead ratios respectively. See Table 2 for details.

In the E665 cryogenic targets the total number of radiation lengths of H and D is low (0.114 and 0.131 respectively), therefore the electron scattered in elastic muon-electron interactions can be reliably reconstructed. It has been shown that the H/D cross-section ratio for elastic muon-electron scattering can be used to check the relative luminosity of the targets [29]. The ratio measured in the current

| Normalization Uncertainty(%) | | | | |
|---|---|---|---|---|
| | Deuterium | Carbon | Calcium | Lead |
| Beam Flux | ±0.48 | ±0.78 | ±0.78 | ±1.06 |
| Areal Target Density | ±0.63 | ±0.02 | ±0.004 | ±0.2 |
| Beam Reconstruction | ±0.61 | ±0.31 | ±0.40 | ±0.5 |
| Total | ±1.00 | ±0.84 | ±0.88 | ±1.19 |
| Cross-Section Ratio | — | 1.30 | 1.33 | 1.55 |

Table 2: Systematic uncertainties in the normalization for cross sections and cross-section ratios.

analysis is consistent with the expected value within 1.9% ± 2.0%.

Another check of the normalization of the deuterium data is provided by the measurement of the cross-section ratio $R^H(x_{Bj}) = \sigma^H(x_{Bj})/(\sigma^D(x_{Bj})/2)$ using the standard procedure of determining $R^A(x_{Bj})$ in this paper. The ratio is found to be independent of $x_{Bj}$ with an average value of $1.00 \pm 0.01$ (stat.) ±0.04 (syst.). This result is consistent with an independent measurement of this ratio within the same experiment using a different data set : $R^H(x_{Bj}) = 1.032 \pm 0.004 \pm 0.017$ [29].

The bin-to-bin systematic uncertainty in $x_{Bj}$ was determined by measuring the sensitivity of the cross-section ratios to the resolution and background removal cuts as well as triggering efficiency. The $y$, $\delta x_{Bj}/x_{Bj}$, $\nu$ and calorimeter cuts were varied independently, and the corresponding variations in the ratios were determined. The largest sensitivity was found in the region $x_{Bj} < 0.0005$, where the systematic uncertainty was as large as 12% for some targets. No uncertainties exceeded 9% in the higher $x_{Bj}$ region. Additional bin-to-bin uncertainties were assigned for hadronic vetoes in the upstream trigger element. The rate of hadronic vetoes in the upstream veto element was found to be as large as 8% in the lowest $x_{Bj}$ bin, but independent of the target type to better than 1% for all $x_{Bj}$. The software simulation of the trigger introduced less than 1% uncertainty. The systematic error due to the $A$ dependence of the fraction of inelastic events removed by the calorimeter cut and residual background contamination by radiative events is estimated to be less than 5%. The values quoted in Table 3 represent typical values of these uncertainties, not the maxima.

The sensitivity to target geometries was estimated by comparing the nominal cross-section ratios with those formed after having eliminated regions of scattered muon phase space where trigger acceptance varied rapidly. The resulting cross-section ratios were consistent to within 4% with those obtained using the full acceptance. The resulting uncertainty has been quoted in Table 3. The bin to bin uncertainties in the ratios were added in quadrature and are displayed as shaded areas in Fig. 2.

Data from a half-thickness carbon target was used to estimate the effects of rescattering in the target and of photon conversions on the measurement of the cross sections. The cross-section ratio of half to full thickness carbon was independent of $x_{Bj}$ to within statistical uncertainties, which were as large as 10%, but indicated that the macroscopic properties of a target did not significantly alter the measurement of the nuclear cross section. The half-thickness carbon data were only used to study systematic effects and are not included in the cross-section ratio analysis.

| Bin-to-Bin Systematic Uncertainties | | | |
|---|---|---|---|
| | $^{12}C/^{2}D$ | $^{40}Ca/^{2}D$ | $^{208}Pb/^{2}D$ |
| $y$ cut | $\sim 2$ % | $\sim 2$ % | $\sim 2$ % |
| $\delta x_{Bj}/x_{Bj}$ cut | $\sim 1$ % | $\sim 1$ % | $\sim 2$ % |
| $\nu$ cut | $\sim 3$ % | $\sim 2$ % | $\sim 3$ % |
| $E_{clus}/\nu$ cut | $\sim 3$ % | $\sim 4$ % | $\sim 5$ % |
| Hadronic Vetoes | $\sim 1$ % | $\sim 1$ % | $\sim 1$ % |
| Trigger Simulation | $\sim 1$ % | $\sim 1$ % | $\sim 1$ % |
| Acceptance | $\sim 3$ % | $\sim 4$ % | $\sim 4$ % |

Table 3: Typical bin-to-bin systematic uncertainties, in the variable $x_{Bj}$, for the cross-section ratios.

# 5 Results

## 5.1 $x_{Bj}$ Dependence

Fig. 2 shows (full dots) the fully corrected, per-nucleon cross-section ratios, $R^A(x_{Bj})$, of carbon, calcium and lead to deuterium using calorimeter background removal. The data indicate that shadowing is present in the low $x_{Bj}$ region in all three nuclei. At very low $x_{Bj}$ the data exhibit little dependence on $x_{Bj}$. From fits of a constant to $R^A(x_{Bj})$, in various $x_{Bj}$ regions, it is found that the shadowing effect is consistent with being independent of $x_{Bj}$ in the regions $x_{Bj} < 0.005$, 0.003 and 0.002, for carbon, calcium and lead, respectively. In addition, the degree of the shadowing is more pronounced in the heavier nuclear targets and is as large as 6%, 15% and 30% in the carbon, calcium and lead nuclei respectively. The open triangle in each plot indicates the ratio, $A_{eff}/A$, measured in photoproduction [2], at fixed photon energy of 60 GeV. The measured values of the cross-section ratios and their errors are compiled in tables 4, 5 and 6.

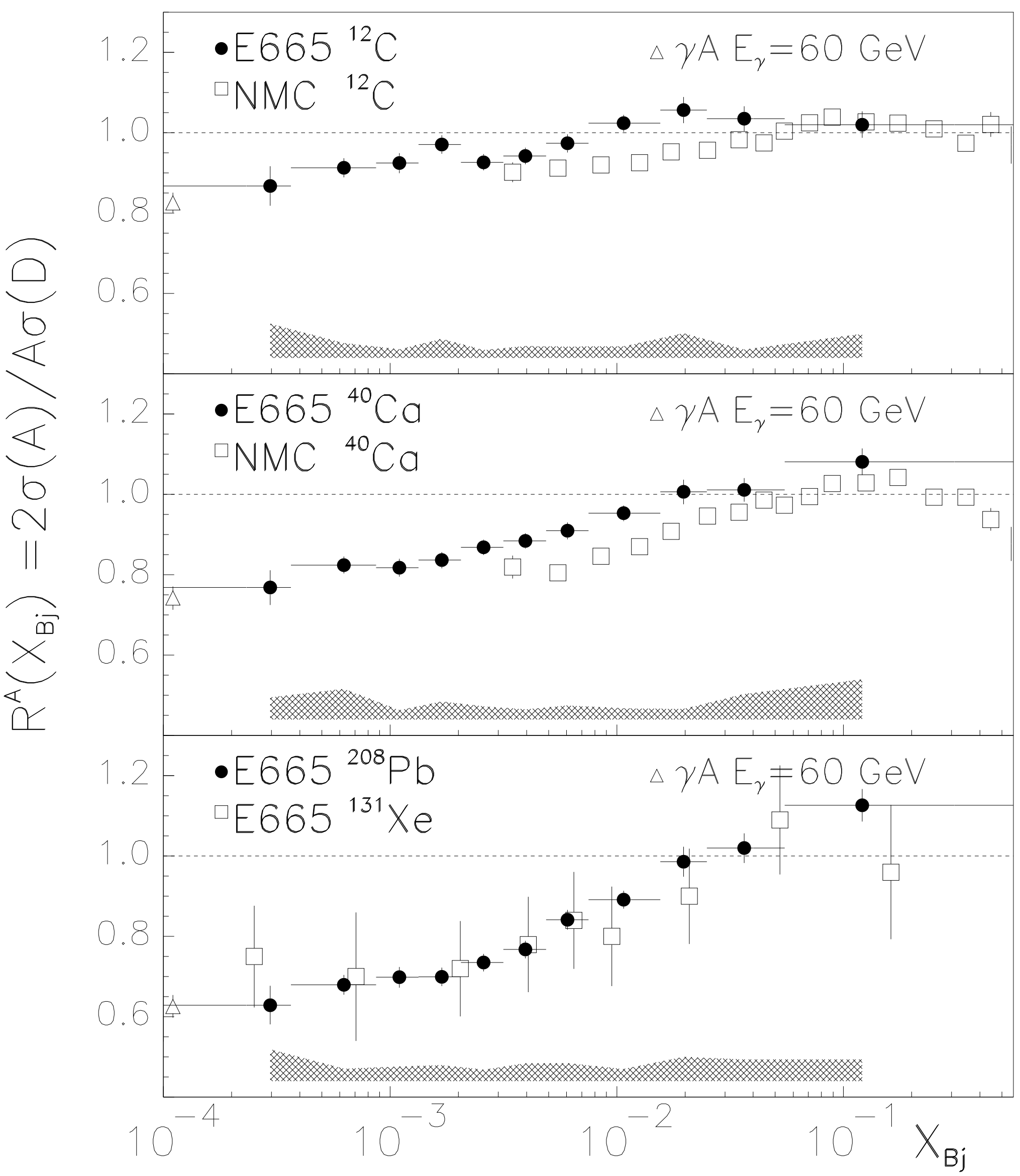

Figure 2: Per-nucleon cross-section ratios, $R^A(x_{Bj})$, for carbon, calcium and lead to deuterium. The shaded band represents the bin-to-bin systematic uncertainty. The overall normalization uncertainties of 1.30%, 1.33% and 1.55%, respectively, have not been included. The vertical error bars represent statistical errors only, horizontal error bars represent the size of the $x_{Bj}$ bin. See text for details.

Included in Fig. 2 are previously published results (empty squares) of the NMC collaboration for carbon and calcium [26] and of the E665 collaboration for xenon [23]. The degree of shadowing measured in this analysis is weaker than that measured by NMC. The difference is smaller when the Ca/C ratios are compared rather than C/D and Ca/D ratios. The NMC data were obtained with a muon beam energy of 280 GeV and were corrected using the FERRAD radiative correction program. The 4−7% systematic shifts of the E665 results with the application of the FERRAD corrections cannot completely account for the difference in the two experiments. It should be noted that at fixed $x_{Bj}$ the average $Q^2$ and $\nu$ are similar in the data samples of E665 and NMC, whereas $< y >$ is lower, by a factor $\sim 2$, in the E665 data.

### 5.2 $Q^2$ and $\nu$ Dependence

The $Q^2$ dependence of $R^A(x_{Bj})$ for each $x_{Bj}$ bin was studied. In each $x_{Bj}$ bin, the cross-section ratio was fit with the function:

$$R^A(x_{Bj}) = a + b \cdot \log_{10}(Q^2/\mathrm{GeV}^2) \quad (\text{ at fixed } x_{Bj}) \tag{3}$$

The slopes, $b$, are shown in Fig. 3 as a function of $x_{Bj}$. Results are consistent with the shadowing effect being independent of $Q^2$, when $x_{Bj}$ is held fixed. The lack of $Q^2$ dependence is consistent with the NMC result (open squares). A similar analysis of the $\nu$ dependence of the shadowing effect was performed. Again, $R^A(x_{Bj})$ was fit to a linear function:

$$R^A(x_{Bj}) = c + d \cdot (\nu/\mathrm{GeV}) \quad (\text{ at fixed } x_{Bj}) \tag{4}$$

In Fig. 4 the fit parameter, $d$, is shown as a function of $x_{Bj}$. The data in Fig. 4 are consistent with no significant $\nu$ dependence in the whole $x_{Bj}$ region considered. From the photoproduction data for $\gamma$Cu in the region $20 < E_\gamma < 185$ GeV [2] a value of $d = -0.0007 \pm 0.0002$ is obtained. The results from the present experiment on $d$, in the lowest $x_{Bj}$ bins where $\nu \approx 210$ GeV, are consistent with the photoproduction result. The fitted values of the parameters $b$ and $d$ are listed in Tables 4, 5 and 6.

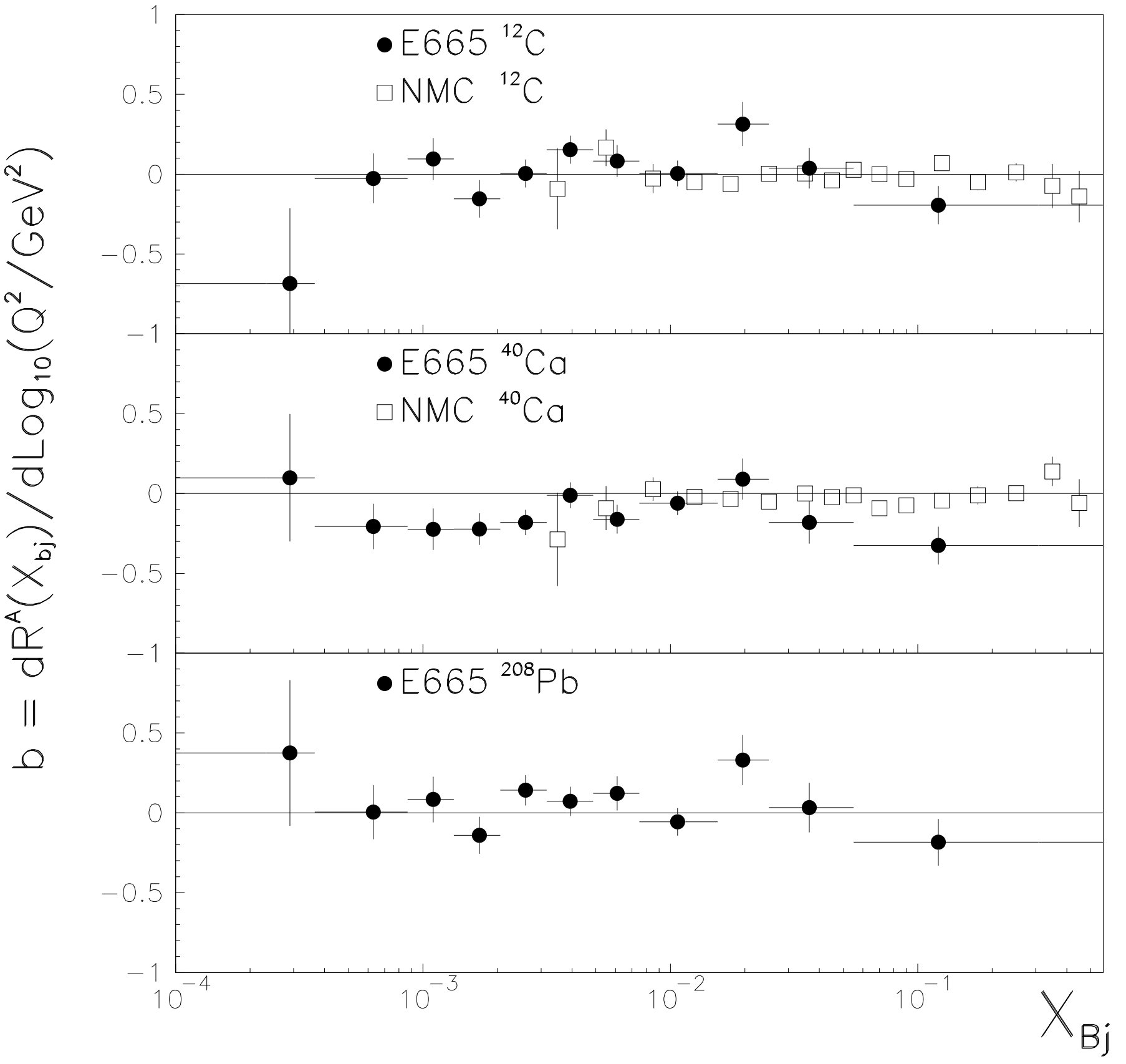

Figure 3: Logarithmic $Q^2$ dependence of the shadowing ratio, $R^A(x_{Bj})$, plotted against $x_{Bj}$. Errors shown are the uncertainty resulting from the fit, in which only statistical errors were considered. The results from [26] are drawn as open squares.

### 5.3 $A$ Dependence

In each $x_{Bj}$ bin, the per-nucleon cross-section ratios were parameterized in terms of the effective number of nucleons, $A^\alpha$, in the nucleus. The strength of the shadowing effect can be expressed by $\alpha$ in the expression $\sigma_{\gamma A} = A^\alpha \sigma_{\gamma N}$, where $\sigma_{\gamma A}$ is the photon-nucleus cross section and $\sigma_{\gamma N}$ is the photon-nucleon cross section, yielding $R^A(x_{Bj}) \propto A^\alpha / A$. In each $x_{Bj}$ bin, $R^A(x_{Bj})$ was fitted with the expression

$$R^A(x_{Bj}) = E_0 \cdot A^{(\alpha-1)} \ \text{( at fixed } x_{Bj}\text{)} \tag{5}$$

The measured value of $\alpha$ for each $x_{Bj}$ bin is shown in Fig. 5 and listed in Table 7. The data indicate that the $A$ dependence of the cross-section ratio is independent of $x_{Bj}$, $\alpha = 0.906 \pm 0.006$, in the region $x_{Bj} \lesssim 0.002$. Fitting the photoproduction data [2] yields $\alpha(\gamma\ A) = 0.904 \pm 0.022$, which is in excellent agrreement with the current analysis. The photproduction result appears in figure 5 as an open triangle.

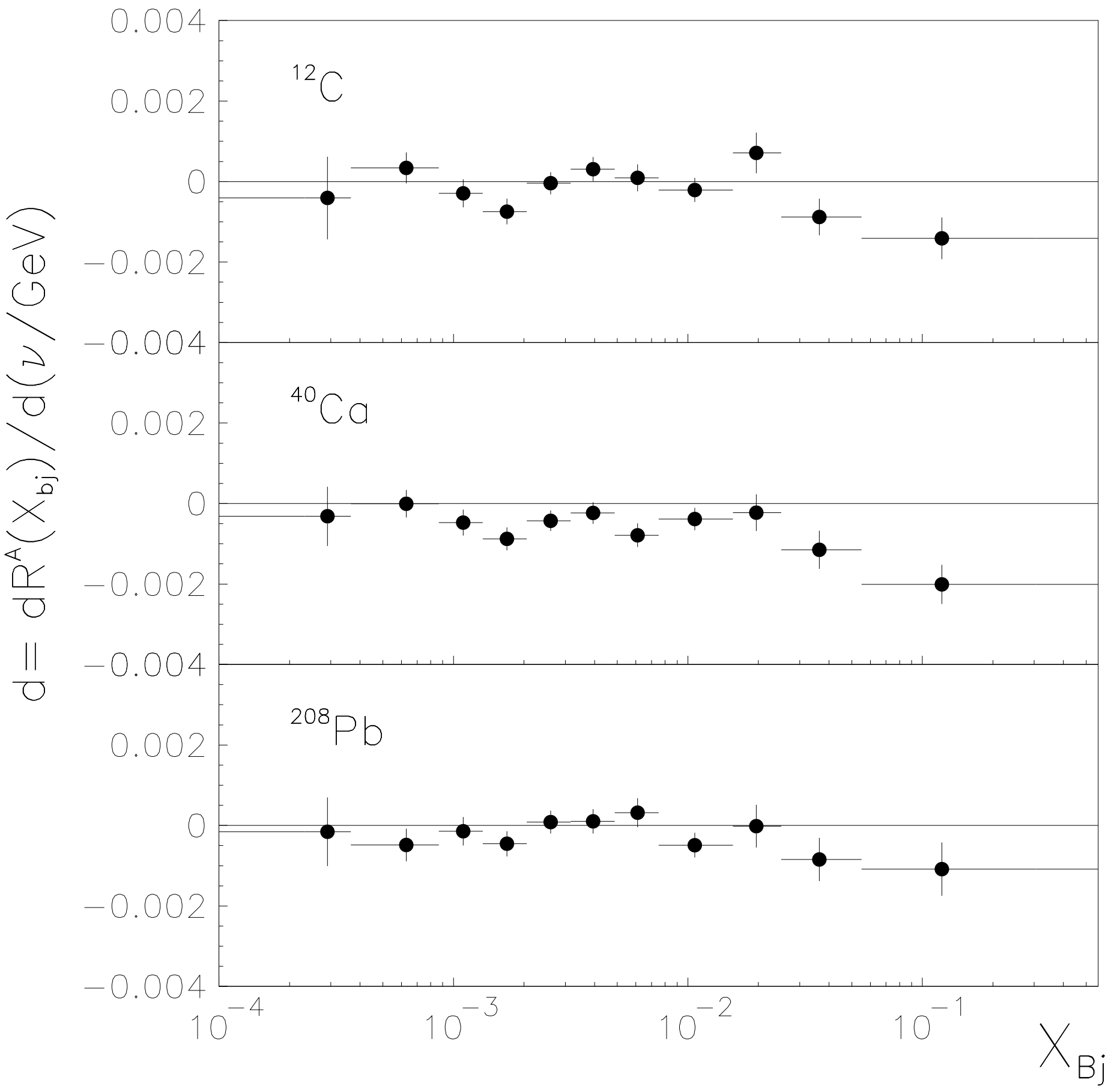

Figure 4: $\nu$ dependence of the shadowing ratio, $R^A(x_{Bj})$, plotted against $x_{Bj}$. Errors shown are the uncertainty resulting from the fit, in which only statistical errors were considered.

In Fig. 6 the $A$ dependence of the shadowing ratio $R^A(x_{Bj})$, as measured in three selected bins of $x_{Bj}$, is compared to that measured in photoproduction ($A_{eff}/A$) [2]. As $x_{Bj}$ decreases, which in the present data sample also implies a decreasing $Q^2$ ( see Table 7), both the size of the shadowing effect as well as its $A$ dependence, at fixed $x_{Bj}$, are approaching the photoproduction limit.

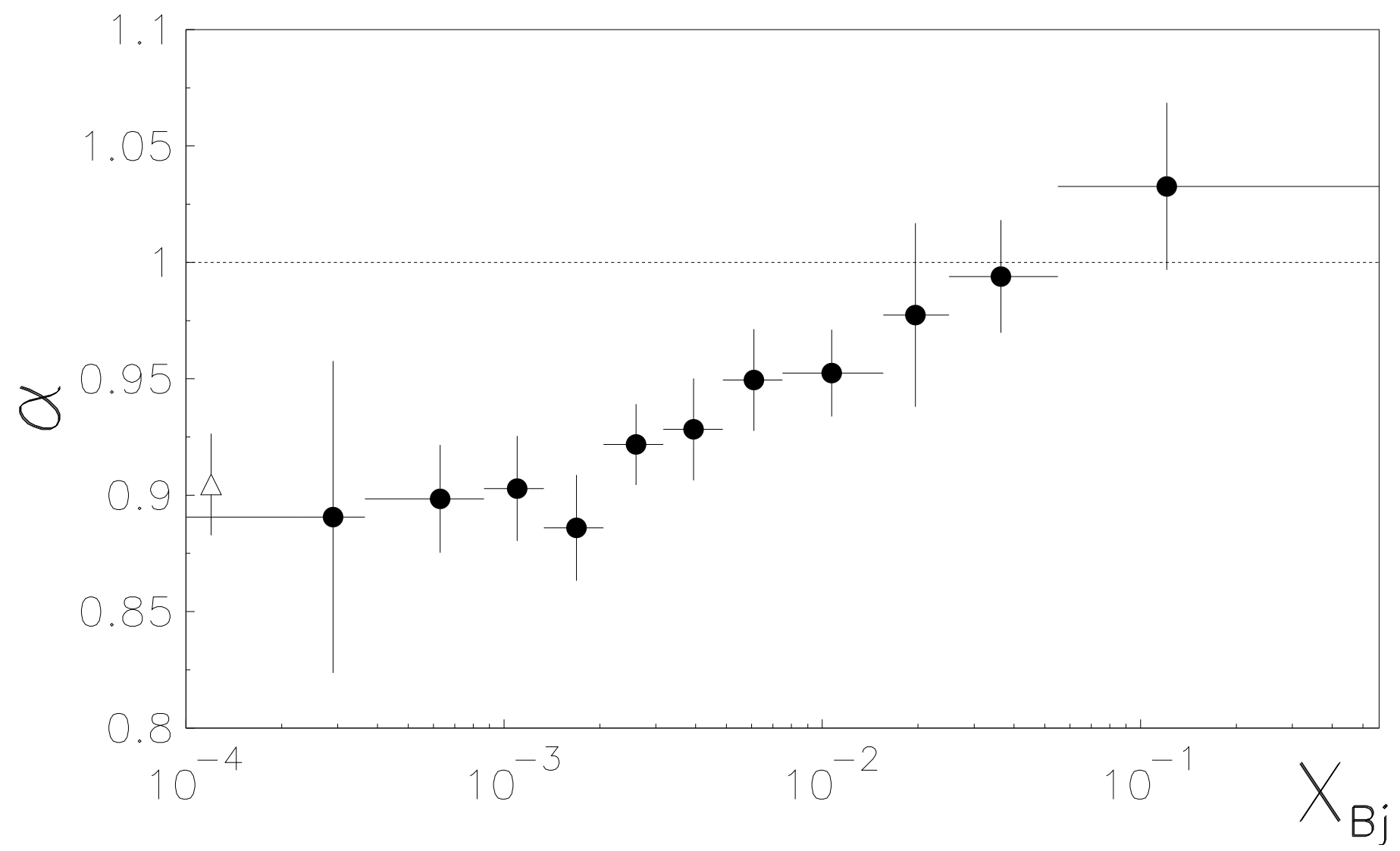

Figure 5: $A$ dependence of shadowing, $R^A(x_{Bj}) \propto A^{\alpha-1}$, for each $x_{Bj}$ bin. Errors shown are the uncertainty resulting from the fit, in which only statistical errors were considered. The open triangle is the corresponding photoproduction point. See text for details.

## 6 Summary

In summary, experiment E665 has measured the per-nucleon cross sections for the scattering of muons on carbon, calcium and lead as compared to that of deuterium. Shadowing has been observed at low $x_{Bj}$ for all three nuclei. No significant $\nu$ or $Q^2$ dependence of shadowing at fixed $x_{Bj}$ is observed. With decreasing $x_{Bj}$ the magnitude of the shadowing effect as well as its $A$-dependence are approaching the photoproduction limit.

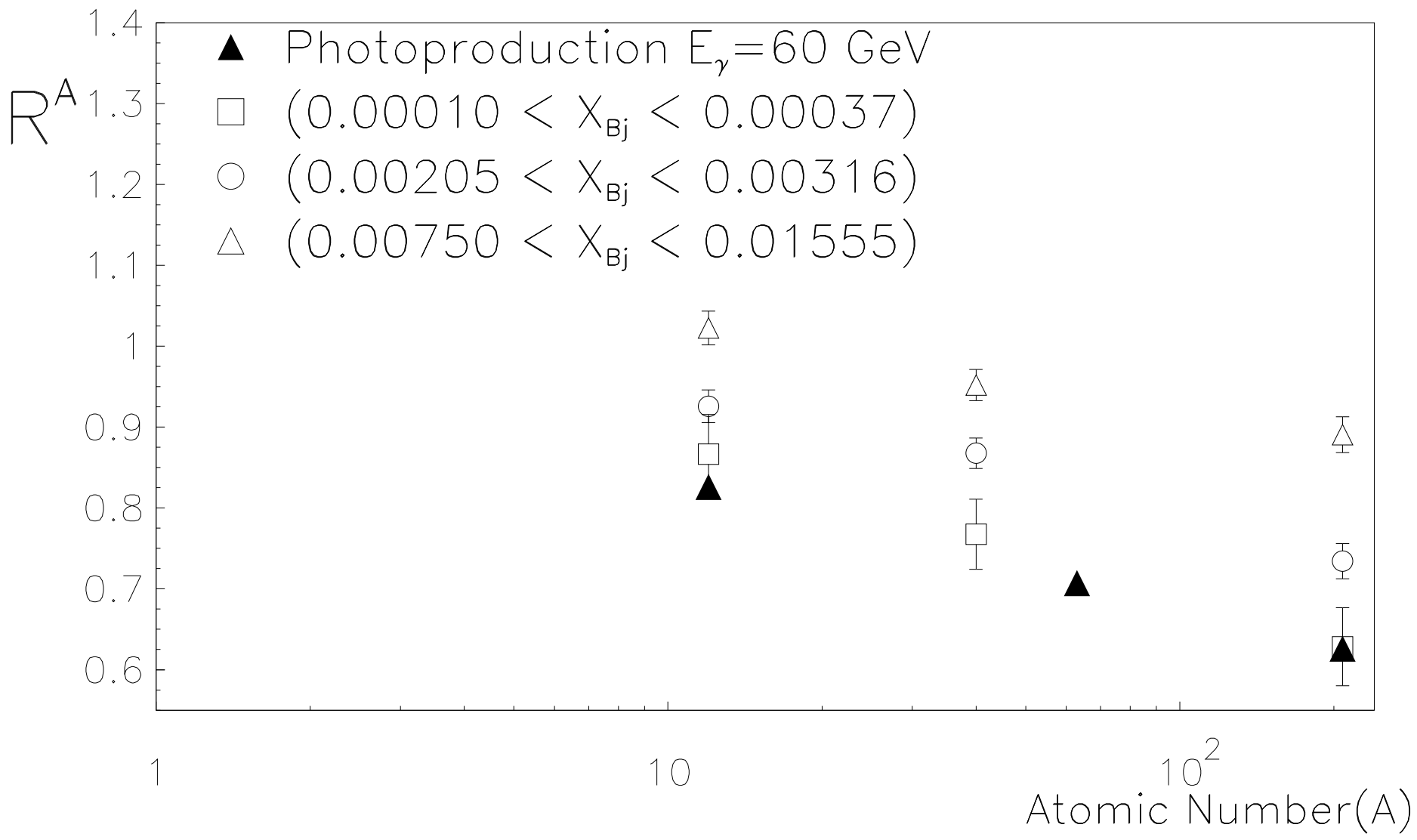

Figure 6: The $A$ dependence of shadowing in three selected $x_{Bj}$ bins, is compared with the $A$ dependence in photoproduction at a photon energy of 60 GeV (solid triangles) [2].

We wish to thank all those personnel, both at Fermilab and at the participating institutions, who have contributed to this experiment. This work was supported by the National Science Foundation, The U.S. Department of Energy, Nuclear Physics and High Energy Physics Divisions, and the Bundesministerium für Forschung und Technologie.

## 7 Tabulated Results

| $^{12}$Carbon | | | | |
|---|---|---|---|---|
| $x_{Bj}$ Range | $R^A(x_{Bj})$ | $R^A(x_{Bj})_{FERR}$ | $b$ | $d \times 10^4$ |
| 0.00010−0.00037 | $0.867 \pm 0.048 \pm 0.085$ | – | $-0.68 \pm 0.47$ | $-4.1 \pm 10$ |
| 0.00037−0.00087 | $0.912 \pm 0.023 \pm 0.037$ | – | $-0.03 \pm 0.16$ | $3.4 \pm 3.8$ |
| 0.00087−0.00133 | $0.924 \pm 0.025 \pm 0.021$ | – | $0.09 \pm 0.13$ | $-2.9 \pm 3.4$ |
| 0.00133−0.00205 | $0.970 \pm 0.023 \pm 0.047$ | – | $-0.15 \pm 0.12$ | $-7.5 \pm 3.2$ |
| 0.00205−0.00316 | $0.926 \pm 0.020 \pm 0.019$ | $0.912 \pm 0.020$ | $0.00 \pm 0.09$ | $-0.4 \pm 2.7$ |
| 0.00316−0.00487 | $0.941 \pm 0.020 \pm 0.029$ | $0.907 \pm 0.019$ | $0.15 \pm 0.09$ | $3.1 \pm 3.0$ |
| 0.00487−0.00750 | $0.973 \pm 0.023 \pm 0.028$ | $0.928 \pm 0.021$ | $0.08 \pm 0.10$ | $0.9 \pm 3.3$ |
| 0.00750−0.01555 | $1.023 \pm 0.021 \pm 0.029$ | $0.959 \pm 0.019$ | $0.00 \pm 0.08$ | $-2.1 \pm 3.0$ |
| 0.01555−0.02500 | $1.055 \pm 0.032 \pm 0.061$ | $1.013 \pm 0.030$ | $0.31 \pm 0.14$ | $7.1 \pm 5.1$ |
| 0.02500−0.05500 | $1.034 \pm 0.030 \pm 0.021$ | $1.016 \pm 0.028$ | $0.04 \pm 0.13$ | $-8.8 \pm 4.5$ |
| 0.05500−0.56234 | $1.019 \pm 0.033 \pm 0.060$ | $1.019 \pm 0.033$ | $-0.19 \pm 0.12$ | $-14 \pm 5.2$ |

Table 4: Per-nucleon cross-section ratio $R^A(x_{Bj})$, and $Q^2$ and $\nu$ dependences as a function of $x_{Bj}$ for carbon. Statistical and bin-to-bin systematic uncertainties are given. A 1.30% overall normalization uncertainty has not been included. The errors on the fit parameters, $b \equiv dR^A/dlog_{10}(Q^2/\mathrm{GeV}^2)$ and $d \equiv dR^A/d(\nu/\mathrm{GeV})$, are those resulting from a fit in which only statistical error were considered. For comparison, $R^A(x_{Bj})_{FERR}$, obtained by applying radiative corrections as calculated by the FERRAD radiative correction program are reported in column 3.

| $^{40}$Calcium | | | | |
|---|---|---|---|---|
| $x_{Bj}$ Range | $R^A(x_{Bj})$ | $R^A(x_{Bj})_{FERR}$ | $b$ | $d \times 10^4$ |
| 0.00010−0.00037 | $0.767 \pm 0.043 \pm 0.055$ | – | 0.10±0.40 | $-3.2 \pm 7.3$ |
| 0.00037−0.00087 | $0.823 \pm 0.021 \pm 0.076$ | – | −0.21±0.14 | $-0.0 \pm 3.4$ |
| 0.00087−0.00133 | $0.817 \pm 0.022 \pm 0.023$ | – | −0.22±0.13 | $-4.7 \pm 3.2$ |
| 0.00133−0.00205 | $0.836 \pm 0.020 \pm 0.045$ | – | −0.22±0.10 | $-8.8 \pm 2.8$ |
| 0.00205−0.00316 | $0.868 \pm 0.019 \pm 0.033$ | $0.823 \pm 0.018$ | −0.18±0.08 | $-4.3 \pm 2.5$ |
| 0.00316−0.00487 | $0.883 \pm 0.019 \pm 0.025$ | $0.850 \pm 0.018$ | −0.01±0.08 | $-2.3 \pm 2.7$ |
| 0.00487−0.00750 | $0.908 \pm 0.021 \pm 0.035$ | $0.842 \pm 0.019$ | −0.16±0.09 | $-7.9 \pm 2.9$ |
| 0.00750−0.01555 | $0.952 \pm 0.019 \pm 0.028$ | $0.889 \pm 0.017$ | −0.06±0.07 | $-3.9 \pm 2.8$ |
| 0.01555−0.02500 | $1.005 \pm 0.031 \pm 0.026$ | $0.957 \pm 0.028$ | 0.09±0.13 | $-2.3 \pm 4.5$ |
| 0.02500−0.05500 | $1.010 \pm 0.030 \pm 0.063$ | $0.946 \pm 0.026$ | −0.18±0.13 | $-11 \pm 4.7$ |
| 0.05500−0.56234 | $1.080 \pm 0.033 \pm 0.100$ | $1.024 \pm 0.032$ | −0.33±0.12 | $-20 \pm 4.8$ |

Table 5: Per-nucleon cross-section ratio $R^A(x_{Bj})$, and $Q^2$ and $\nu$ dependences as a function of $x_{Bj}$ for calcium. Statistical and bin-to-bin systematic uncertainties are given. A 1.33% overall normalization uncertainty has not been included. The errors on the fit parameters, $b \equiv dR^A/dlog_{10}(Q^2/\mathrm{GeV}^2)$ and $d \equiv dR^A/d(\nu/\mathrm{GeV})$, are those resulting from a fit in which only statistical error were considered. For comparison, $R^A(x_{Bj})_{FERR}$, obtained by applying radiative corrections as calculated by the FERRAD radiative correction program are reported in column 3 .

| $^{208}$Lead | | | | |
|---|---|---|---|---|
| $x_{Bj}$ Range | $R^A(x_{Bj})$ | $R^A(x_{Bj})_{FERR}$ | $b$ | $d \times 10^4$ |
| 0.00010−0.00037 | $0.628 \pm 0.048 \pm 0.079$ | – | 0.38±0.47 | $-1.6 \pm 8.5$ |
| 0.00037−0.00087 | $0.679 \pm 0.025 \pm 0.031$ | – | 0.00±0.17 | $-4.9 \pm 4.0$ |
| 0.00087−0.00133 | $0.698 \pm 0.026 \pm 0.036$ | – | 0.08±0.14 | $-1.4 \pm 3.5$ |
| 0.00133−0.00205 | $0.699 \pm 0.023 \pm 0.040$ | – | −0.14±0.12 | $-4.5 \pm 3.1$ |
| 0.00205−0.00316 | $0.734 \pm 0.022 \pm 0.029$ | $0.704 \pm 0.020$ | 0.14±0.10 | $0.9 \pm 2.8$ |
| 0.00316−0.00487 | $0.767 \pm 0.022 \pm 0.044$ | $0.730 \pm 0.020$ | 0.07±0.09 | $1.0 \pm 3.0$ |
| 0.00487−0.00750 | $0.841 \pm 0.025 \pm 0.049$ | $0.778 \pm 0.022$ | 0.12±0.11 | $-3.2 \pm 3.6$ |
| 0.00750−0.01555 | $0.891 \pm 0.022 \pm 0.030$ | $0.850 \pm 0.021$ | −0.06±0.09 | $-4.9 \pm 3.1$ |
| 0.01555−0.02500 | $0.985 \pm 0.037 \pm 0.061$ | $0.897 \pm 0.033$ | 0.33±0.16 | $-0.2 \pm 5.3$ |
| 0.02500−0.05500 | $1.019 \pm 0.037 \pm 0.054$ | $0.951 \pm 0.032$ | 0.03±0.16 | $-8.4 \pm 5.3$ |
| 0.05500−0.56234 | $1.125 \pm 0.040 \pm 0.054$ | $1.046 \pm 0.039$ | −0.18±0.15 | $-10 \pm 6.6$ |

Table 6: Per-nucleon cross-section ratio $R^A(x_{Bj})$, and $Q^2$ and $\nu$ dependences as a function of $x_{Bj}$ for lead. Statistical and bin-to-bin systematic uncertainties are given. A 1.55% overall normalization uncertainty has not been included. The errors on the fit parameters, $b \equiv dR^A/dlog_{10}(Q^2/\mathrm{GeV}^2)$ and $d \equiv dR^A/d(\nu/\mathrm{GeV})$, are those resulting from a fit in which only statistical error were considered. For comparison, $R^A(x_{Bj})_{FERR}$, obtained by applying radiative corrections as calculated by the FERRAD radiative correction program are reported in column 3.

| $x_{Bj}$ Range | $< Q^2 >$ (GeV$^2$) | $< y >$ | $\alpha$ |
|---|---|---|---|
| 0.00010−0.00037 | 0.15 | 0.59 | 0.891±0.067 |
| 0.00037−0.00087 | 0.26 | 0.49 | 0.899±0.023 |
| 0.00087−0.00133 | 0.39 | 0.41 | 0.903±0.023 |
| 0.00133−0.00205 | 0.51 | 0.34 | 0.886±0.023 |
| 0.00205−0.00316 | 0.67 | 0.27 | 0.922±0.017 |
| 0.00316−0.00487 | 0.93 | 0.29 | 0.928±0.022 |
| 0.00487−0.00750 | 1.35 | 0.25 | 0.950±0.022 |
| 0.00750−0.01555 | 2.42 | 0.25 | 0.952±0.019 |
| 0.01555−0.02500 | 4.45 | 0.26 | 0.977±0.039 |
| 0.02500−0.05500 | 7.91 | 0.24 | 0.993±0.024 |
| 0.05500−0.56234 | 22.5 | 0.22 | 1.033±0.036 |

Table 7: Average values of $Q^2$ and $y$, for each $x_{Bj}$ bin, and the $A$ dependence of shadowing, $R^A(x_{Bj}) \approx A^{\alpha-1}$, in bins of $x_{Bj}$. Errors on $\alpha$ are those resulting from the fit.